\documentclass[twocolumn,superscriptaddress,a4paper,pra,showpacs,floatfix] {revtex4-1}
\usepackage{dsfont}
\usepackage{pifont}
\usepackage[ngerman,english]{babel}
\usepackage{natbib}
\usepackage{graphicx}
\usepackage[FIGTOPCAP]{subfigure}
\usepackage{amsmath}
\usepackage{tikz}
\usepackage[breaklinks,colorlinks,allcolors=blue]{hyperref}
\usepackage{amsfonts}
\usepackage{color}
\usepackage[section]{placeins}

\newcommand*\circled[1]{\raisebox{.5pt}{\textcircled{\raisebox{-.9pt} {#1}}}}

\makeatletter
\newcommand*{\rom}[1]{\expandafter\@slowromancap\romannumeral #1@}
\makeatother

\newcommand{\pp}[2]{\frac{\partial #1}{\partial #2}}

\begin{document}

\title{Attosecond dynamics of light-induced resonant hole transfer in high-order-harmonic generation}
\author{Jhih-An You}
\email{jhihan@gmail.com}
\affiliation{Max Planck Institute for the Structure and Dynamics of Matter,Luruper Chaussee 149, D-22761 Hamburg, Germany}
\affiliation{Center for Free-Electron Laser Science, Luruper Chaussee 149, D-22761 Hamburg, Germany}
\author{Jan Marcus Dahlstr\"{o}m}
\affiliation{Department of Physics, Stockholm University, AlbaNova University Center, SE-106 91 Stockholm, Sweden}
\author{Nina Rohringer}
\email{nina.rohringer@mpsd.mpg.de}
\affiliation{Max Planck Institute for the Structure and Dynamics of Matter,Luruper Chaussee 149, 22761 Hamburg, Germany}
\affiliation{Center for Free-Electron Laser Science, Luruper Chaussee 149, 22761 Hamburg, Germany}

\begin{abstract}
We present a study of high-order-harmonic generation (HHG) assisted by extreme ultraviolet (XUV) attosecond pulses, which can lead to the excitation of inner-shell electrons and the generation of a second HHG plateau. With the treatment of a one-dimensional model of krypton, based on time-dependent configuration interaction singles (TDCIS) of an effective two-electron system, we show that the XUV-assisted HHG spectrum reveals the duration of the semiclassical electron trajectories. The results are interpreted by the strong-field approximation (SFA) and the importance of the hole transfer during the tunneling process is emphasized. Finally, coherent population transfer between the inner and outer holes with attosecond pulse trains is discussed.
\end{abstract}

\maketitle

\section{Introduction}
High-order-harmonic generation (HHG) is a fascinating nonperturbative phenomenon where high-frequency photons are produced through the interaction of a low-frequency intense laser field with atomic or molecular gases \cite{Corkum1993,Lewenstein1994a,Krausz2009}. The HHG spectrum forms a plateau in the XUV or soft-x-ray region \cite{Popmintchev2012}, that ends abruptly at a cutoff energy $3.17U_p+1.32I_p$, where $I_p$ is the ionization potential and $U_p$ is the ponderomotive energy of the electron in the laser field \cite{Corkum1993}. HHG can be explained by a semiclassical three-step model, in which the intense laser field distorts the atomic potential such that the valence electron is tunnel ionized, accelerated in the continuum and then recombined to the ground state with an emitted photon. The broadband property of the plateau supports attosecond pulse generation so that the time-dependent measurement on the attosecond time scale become possible \cite{Krausz2009,Nisoli2009,Sansone2011,Mairesse2004,Mairesse2003b,Popmintchev2012}. In addition, the harmonics generated from the recombination process contain information about the interference between the ground state and the continuum wavepacket, and this interference has many applications for the studies of molecular structure and dynamics \cite{Niikura2002,Kanai2005,Lein2002,Baker2006,Kanai2007,Midorikawa2011,Shiner2011,Kraus2013,Zhang2015}. With the progress of laser technology both harmonic intensity and photon energy will increase, allowing for controlling strong optical and attosecond-pulse driven electron dynamics.\par

One way to manipulate the HHG process is by applying multi-color laser fields that provide various ways to change the HHG spectrum. 
Additional frequencies in combination with the driving laser field can alter the semiclassical electron trajectories in the continuum, which can give rise to change HHG intensity by constructive interference of different orbitals.
This idea can be used to enhance the HHG plateau \cite{Kondo1996,Brizuela2013} or extend the cutoff energy \cite{Mansten2008,Ivanov2009,Chipperfield2009}. Another approach to enhance the HHG is to control the time of ionization by using attosecond pulse trains to initialize the ionization time via single photon absorption \cite{Schafer2004,Gaarde2005,Takahashi2007,Brizuela2013}. Furthermore, the temporal property of the generated attosecond pulses can be controlled by varying the phase difference between the laser field and a second harmonic field \cite{Mauritsson2006,MashikoPRL2008}. If the second harmonic field is weak it will barely modify the original odd harmonics, but the symmetry breaking between half optical cycles leads to the generation of weak even harmonics. These even harmonics have been used to retrieve emission times of attosecond pulses and recombination times of the electrons \cite{Dudovich2006,Dahlstrom2011,Shafir2012,ZhaoPRL2013}. 
Some ideas beyond single-active electron approximation have also been theoretically discussed to extend the HHG cut-off. 
For example, nonsequential double recombination,
which is the inverse of single-photon double ionization, leads to a second plateau, but 
the probability for this process was found to be extremely small \cite{Koval2007}. 
Here, we will study a two-electron scheme to generate a second plateau by inner-shell excitation with an assisting XUV pulses, as first proposed by Buth \textit{et al}. \cite{Buth2011,Buth2012,Kohler2012,Buth2015}: 
During the excursion of a valence electron in continuum, one can excite an inner-shell electron to the vacant valence state. Then, the returning electron recombines into the inner-shell hole, leading to an increase in energy of the emitted photon. Besides the creation of a second HHG plateau, this process also reveals detailed information about the continuum trajectories of the electrons participating in the HHG process.\par

In this paper, we study resonant XUV-assisted HHG with the time-dependent-configuration-interaction singles (TDCIS) approach of an effective two-electron system to model the krypton atom with $3d$ and $4p$ orbitals.
We consider two different cases -- few cycle IR pulse + single XUV attopulse and flat-top IR field + XUV attosecond pulse train (APT). In both cases, the HHG process can be controlled by varying the time delay (or phase delay in the case of IR-APT set-up) of IR and XUV field. A semiclassical model and stationary phase approximation beyond the strong field approximation (SFA) is presented to interpret the results, where the XUV field is treated as a perturbation to explain the hole dynamics on the subcycle time scale \cite{Buth2012,Buth2011,Kohler2012,Buth2015}. In the few cycle IR pulse + single XUV attopulse case, the temporal information of electron trajectories is exhibited. Our studies also reveal discrepancies between the TDCIS calculation and the SFA model. We propose a modification of the stationary phase approximation in the extended SFA model and provide an understanding of hole transition during the tunneling. In the flap-top IR field + XUV PT case, the effect of the repetition of XUV pulses on the inner-shell population is also discussed.
\par

This paper is organized as follows:  
In Sec. \ref{Theoretical_method}, we present our theoretical methods: time-dependent configuration interaction and the semiclassical model, which is a generalization of the Lewenstein's model \cite{Ivanov1996,Lewenstein1994a}. 
In Sec. \ref{Result_and_Discussion}, we compare different HHG spectra with respect to different time delay for both flat-top and few-cycle IR pulses. We analyze the discrepancy between different models. Finally, Sec. \ref{Conclusion} contains our  conclusions and outlook. 

\section{Theoretical method}  \label{Theoretical_method}

In order to model XUV-assisted HHG we need at least two electrons and the possibility for stimulated transition between different electronic shells. We consider a two-electron one-dimensional (1D) model system: 
\begin{equation} \label{2eTDSE}
H(t)=T_1+T_2+V(z_1)+V(z_2)+V_{ee}(z_1,z_2)+V_{\text{ext}}(t),
\end{equation}
where $T_i$ is the kinetic operator of electron $i$, $V(z_i)=-{Z_{\text{eff}}}/\sqrt{{z_i}^2+r_c^2}$ is an effective atomic potential \cite{Schwengelbeck1994,Javanainen1988}, $V_{ee}(z_1,z_2)=1/\sqrt{(z_1-z_2)+r_{ee}^2}$ is the 1D Coulomb interaction between the electrons 
and $V_{\text{ext}}(t)=[E_\text{IR}(t)+E_\text{XUV}(t)](z_1+z_2)$ is the external potential due to the laser-electron interaction within the dipole approximation. Here $r_c,r_{ee}$ and $Z_{\text{eff}}$ are determined to reproduce the binding potential and the 3d-4p excitation energy of krypton. 
\subsection{Time-dependent configuration interaction}
Although solving the time-dependent Schr\"{o}dinger equation is possible for a two-electron system, the computation time increases dramatically in the strong-field regime.
A time-dependent configuration interaction singles approach provides an efficient and sufficient \footnote{We have compared full time-dependent solution of the problem to TDCIS solutions and did not find any discrepancies between both methods} treatment of different many-body effects and coupled channels \cite{Rohringer2006}. In TDCIS formalism, we consider the spin-triplet two-electron Hartree-Fock ground state $|\Phi_0 \rangle$ and its single excitations $|\Phi_i^a \rangle$ based on the one-particle Fock operator $\hat{H_F}$ and its eigenstate $|\varphi_p\rangle$ with energy $\epsilon_p$.
Here, indices $i$,$j$,$k$,$l$,... are used for spatial orbitals that are occupied in $|\Phi_0 \rangle$, and the indices $a$,$b$,$c$,... are initially unoccupied (virtual) orbitals. 
The ground state is chosen as a spin-triplet state such that the two active electrons do not fill the same orbital.
In the following, we neglect the spin because the electric dipole transition does not change the spin configuration.
The many-body wave packet in the TDCIS approximation is expanded by
\begin{equation} \label{wave_packet}
|\Psi,t \rangle=\alpha_0(t)|\Phi_0 \rangle+\sum_{i}\sum_{a}\alpha_i^a(t)|\Phi_i^a \rangle,
\end{equation}
with the initial conditions $\alpha_0(t_0)=1$ and $\alpha_i^a(t_0)=0$.
To understand the hole dynamics and the corresponding wavepacket propagating in real space, 
we introduce a time-dependent orbital \cite{Rohringer2006} that collects all the excitations originating from the occupied orbital $|\varphi_i\rangle$,
\begin{equation}
|\chi_i,t\rangle=\sum_{a}\alpha_i^{a}(t)|\varphi_a\rangle.
\end{equation}
For the atomic system interacting with laser field $E(t)$ linearly polarized along the $z$ axis, the TDCIS equations of motion can be written as \cite{Rohringer2006}
\begin{equation}  \label{TDCIS_ground}
i\dot{\alpha_0}=-E(t)\sum_i \langle\varphi_i|\hat{z}|\chi_i,t\rangle
\end{equation}
\begin{align}  %
i\frac{\partial}{\partial t}|\chi_i\rangle=&\overbrace{(\hat{H}_F-\varepsilon_i)|\chi_i\rangle}^{\circled{1}} + \overbrace{\sum_{i'} \hat{P}\{\hat{K}_{i'i}-\hat{J}_{i'i}\}|\chi_{i'}\rangle}^{\circled{2}}
\label{TDCIS_nodecay}\\  
&\underbrace{-E(t)\hat{P}\hat{z}\{\alpha_0 |\varphi_i \rangle  +|\chi_i \rangle \}}_{\circled{3}}+\underbrace{E(t)\sum_{i'}z_{ii'}|\chi_{i'} \rangle}_{\circled{4}},\nonumber
\end{align}
where $z_{ii'}=\langle\phi_i|\hat{z}|\phi_{i'}\rangle$ is the dipole transition matrix element, $\hat{P}$ is the projection operator acting on the subspace composed of the virtual orbitals
\begin{equation}
\hat{P}=\sum_a^{\text{virt}} |\phi_a\rangle\langle\phi_a|=\mathbb{I}-\sum_i^{\text{occ}}|\phi_i\rangle\langle\phi_i|, 
\end{equation}
and $\hat{J}_{i'i}$ and $\hat{K}_{i'i}$ are, respectively, generalized Coulomb and exchange operators defined with the direct Coulomb matrix elements $v_{ai'a'i}$ and the exchange matrix elements $v_{ai'ia'}$:
\begin{equation}
\begin{split}
\langle \varphi_a |\hat{J}_{i'i}| \varphi_{a'} \rangle &= v_{ai'a'i} \\
\langle \varphi_a |\hat{K}_{i'i}| \varphi_{a'} \rangle &= v_{ai'ia'}.
\end{split}
\end{equation}
This procedure establishes a system of linear, coupled one-particle Schr\"{o}dinger-like equations in Eq.\ \eqref{TDCIS_nodecay} for the orbitals $|\chi_i,t\rangle$ with initial condition $|\chi_i,t_0\rangle=0$. Different kinds of coupling are separated in different terms in Eq.\ (\ref{TDCIS_ground}) and (\ref{TDCIS_nodecay}). For example,  term $\circled{2}$ in Eq.\ (\ref{TDCIS_nodecay}) represents the channel-coupling due to electron-electron correlation, while term $\circled{4}$ represents the laser-driven channel-coupling between different ionic states. The transition between the ground state and the electron-hole wavepacket is contained in the left part of term $\circled{3}$ in Eq.\ (\ref{TDCIS_nodecay}) and in Eq.\ (\ref{TDCIS_ground}). The contribution of a particular pathway can easily be examined by removing the associated coupling terms. However, the TDCIS formalism does not allow for double excitations that may occur by the full Hamiltonian of Eq.~(\ref{2eTDSE}). \par

All information, including the physical observables, can be obtained by calculating $\alpha_0$ and $| \chi_i\rangle$.
For example, the expectation value of a one-body operator $\hat{D}$, such as a dipole operator or a dipole acceleration operator, can be expressed as:
\begin{align}
\langle \Psi,t|\hat{D}|\Psi,t\rangle= & |\alpha_0|^2\sum_{i}d_{ii}+\sum_{i}\langle \chi_i,t| \hat{d}|\chi_i,t\rangle \nonumber\\
& + \sum_{i}d_{ii}\sum_{j}\langle \chi_j,t| \chi_j,t\rangle - \sum_{ii'}d_{ii'}\langle \chi_i,t| \chi_{i'},t\rangle \nonumber\\
& +2\mbox{ Re}\left(\alpha_0\sum_{i} \langle \chi_i,t| \hat{d}|\varphi_i\rangle \right),
\end{align}
where $\hat{d}$ is the related single-particle operator with matrix element $d_{ii'}=\langle\phi_i|\hat{d}|\phi_{i'}\rangle$.
We can obtain the HHG power spectrum by Fourier-transforming the time dependent dipole in length gauge $d_l(t)$ or dipole acceleration $d_a(t)$
\begin{equation}\label{FTd}
P(\Omega)=\bigg|\int_{-\infty}^{\infty} dt e^{i\Omega t} d_a(t)\bigg|^2=\bigg|\Omega^2 \int_{-\infty}^{\infty} dt e^{i\Omega t} d_l(t)\bigg|^2.
\end{equation}
To unravel the dynamics of ionic states, we can construct the reduced density matrix of the residual ion by tracing over the virtual orbital $a$ \cite{Blum1996,Rohringer2006,Rohringer2009a}: 
\begin{equation}
	\rho^{( \text{ion} )}_{ij}(t)\equiv \sum_a \alpha_i^{a*}(t)\alpha_j^{a}(t) = \langle \chi_i,t| \chi_{j},t\rangle,
\end{equation}
The diagonal term $\rho^{( \text{ion} )}_{ii}$ is the probability of forming a hole in orbital $|\varphi_i \rangle$, or in other words, forming an ion from orbital $|\varphi_i \rangle$. 
In this work, we will use the notation $\rho_{1}(t)$ and $\rho_{2}(t)$ as the inner hole population and the outer hole population, respectively.
\subsection{Semiclassical model} \label{Semi-classical Model}
To get more physical insight we also consider a semiclassical model of XUV-assisted HHG. The XUV field is treated as a perturbation in connection with Lewenstein's semiclassical calculation for an atom in an intense, low-frequency field. The two-electron Schr\"{o}dinger equation takes the form,
\begin{equation}
 [\hat{H}_0+\hat{H}_\text{IR}(t)+\lambda \hat{H}_\text{X}(t)]| \Phi (t) \rangle=i\frac{\partial}{\partial t}| \Phi (t) \rangle,
\end{equation}
where $\hat{H}_0=\hat{h}_{0}\otimes\hat{\mathds{1}}+\hat{\mathds{1}}\otimes\hat{h}_{0}$ represents the atomic electronic structure,
$\hat{H}_\text{IR}=\hat{h}_\text{IR}\otimes\hat{\mathds{1}}+\hat{\mathds{1}}\otimes\hat{h}_\text{IR}=E_\text{IR}(t)\hat{z}_1+E_\text{IR}(t)\hat{z}_2$ is the interaction with the optical laser, 
and $\hat{H}_\text{X}=\hat{h}_\text{X}\otimes\hat{\mathds{1}}+\hat{\mathds{1}}\otimes\hat{h}_\text{X}=E_\text{XUV}(t)\hat{z}_1+E_\text{XUV}(t)\hat{z}_2$ is the interaction with an XUV field that is treated as a perturbation. 
Here we use capital letters to represent two-particle operators and small letters to represent one-particle operators. 
In this semiclassical model, we consider uncorrelated dynamics and represent the wave function as a Slater determinant. 
The spatial one-electron states are represented as the core state $|1\rangle$ with energy $\varepsilon_1$, valence state $|2\rangle$ with energy $\varepsilon_2$ (ionization potential $I_p=-\varepsilon_2$), and the continuum states with canonical momentum $p$ and vector potential $A(t)$ associated with $E_{\text{IR}}$ at time $t$ is $|k(t)\rangle=|p+A(t)\rangle$. The related transition matrix elements are $\langle2|z|1\rangle=z_{12}$, $\langle k|z|1\rangle=d_{1}(k)$, and $\langle k|z|2\rangle=d_{2}(k)$. The relevant two-particle state can be written as
the ground state $| 1,2 \rangle=2^{-1/2}[| 1 \rangle\otimes| 2 \rangle-| 2 \rangle\otimes| 1 \rangle]$ with energy $E_0=\varepsilon_1+\varepsilon_2$, the states with one electron in the continuum $k$ and the other one in the inner shell 
$| 1,k \rangle=2^{-1/2}[| 1 \rangle\otimes| k \rangle-| k \rangle\otimes| 1 \rangle]$, and the states with the electron in the continuum $k$ and the other one in the outer shell
$| 2,k \rangle=2^{-1/2}[| 2 \rangle\otimes| k \rangle-| k \rangle\otimes| 2 \rangle]$.\par
To calculate the time-dependent wave function, we treat the XUV interaction as a perturbation and expand the wave function into the unperturbed part $| \Phi_\text{IR} (t) \rangle$ and the perturbed part $| \Phi_\text{X} (t) \rangle$:
\begin{equation}
 | \Phi (t) \rangle=| \Phi_\text{IR} (t) \rangle+\lambda| \Phi_\text{X} (t) \rangle.
\end{equation}
Here $| \Phi_\text{IR} (t) \rangle$ satisfies the Schr\"{o}dinger equation without the XUV field
\begin{equation}
 i\frac{\partial}{\partial t}| \Phi_\text{IR} (t) \rangle= [\hat{H}_0+\hat{H}_\text{IR}(t)]| \Phi_\text{IR} (t) \rangle.
\end{equation}
Based on the SFA and the derivation from Lewenstein's model \cite{Lewenstein1994a}, the solution of $\Phi_\text{IR}$ can be written as $| \Phi_\text{IR} (t) \rangle= | \Phi_g (t) \rangle + | \Phi_c (t) \rangle$, where $|\Phi_g (t)\rangle=e^{-iE_0(t-t_0)}| 1,2 \rangle$ is the time-dependent ground-state wave function and 
\begin{equation} \label{unperturbed}
 | \Phi_c (t) \rangle \equiv \int a_{k}(t)| 1,k \rangle dk
\end{equation}
is the time-dependent continuum state wave function with
\begin{equation}
 a_{k}(t) = -i\int_{t_0}^{t} dt_1 e^{-i\tilde{S}_{1}(t,t_1,p)}  \langle p+A(t_1)|\hat{h}_\text{IR}(t_1)| 2 \rangle,
\end{equation}
where $p=k-A(t)$ is the canonical momentum and is a conserved quantity in SFA.
Here
\begin{equation}
\tilde{S}_{1}(t,t_1,p)=\int^t_{t_1}dt'\frac{1}{2}[p+A(t')]^2+\varepsilon_1(t-t_1)+E_0(t_1-t_0)
\end{equation}
is the semiclassical action.
To get the HHG spectrum, we need to calculate the time-dependent dipole expectation, neglecting of the c-c part: \cite{Lewenstein1994a}
$\langle \Phi (t)|z| \Phi (t) \rangle\approx d_{gc}(t) + \text{c.c.}$, where $d_{gc}(t)=\langle \Phi_g (t)|z| \Phi_c (t) \rangle$. 
During the integration over $t_1$ and $k$, the dipole moment $d_{gc}(t)$ can be simplified by using stationary phase approximation (SPA) separately due to the fast oscillation of $\tilde{S}_{1}$, and this approach allows a straightforward connection to classical trajectories. 
First, the contribution from the saddle point $p=p_{(s)}$ of  $S_1(t,t_1,p)=\tilde{S}_{1}(t,t_1,p)-E_0(t-t_0)=\int_{t_1}^t dt'\left\{[p-A(t')]^2/2+I_p\right\}$ results in the classical recollision condition \cite{Lewenstein1994a} 
\begin{equation} \label{momentum}
p_{(s)} (t,t_1)=-\frac{\int_{t_1}^t A(t') dt'}{t-t_1},
\end{equation}
and we can get
\begin{equation}\label{IRdipole}
\begin{split}
d_{gc}(t)\approx & -i\int_{t_0}^t dt_1 \text{d}_2(p_{(s)}+A(t_1))E_{\text{IR}}(t_1) \\ &\times a_\text{pr}(t,t_1)a_\text{rec}(t,t_1).
\end{split}
\end{equation}
Here
\begin{equation} \label{dipole_coefficient_IR}
\begin{split}
a_\text{pr}(t,t_1)&=\bigg(\frac{2\pi i}{t-t_{1}}\bigg)^{1/2}e^{-iS_1(t,t_1,p_{(s)})} \\
a_\text{rec}(t,t_1)&=\text{d}_2^*(p_{(s)}+A(t))
\end{split}
\end{equation}
represent the amplitudes of the propagation and the recombination process.
Then the integral can be furthermore factorized into several terms (see the appendix)
\begin{equation} \label{IR_factorization}
d_{gc}(t) \approx \sum_{t_i}\frac{1}{\sqrt{i}}[a_\text{ion}(t,t_i)a_\text{pr}(t,t_i)a_\text{rec}(t,t_i)]
\end{equation}
 by implementing the SPA around $t_1=t_{1(s)}$, which is shifted to the complex plane from the real ionization time $t_i$ defined as $p_{(s)}+A(t_i)=0$ according to the three-step model \cite{Corkum1993,Ivanov1996}.
Here $a_\text{ion}$ is the amplitude of the tunneling ionization process originated from the integration in the semiclassical action along the imaginary direction $S_1(t_i,t_{1(s)},p_{(s)})$, while $S_1(t,t_i,p_{(s)})$ is shown in $a_\text{pr}(t,t_i)$. More details are given in the appendix. For each $t$, the time-dependent dipole is determined by the discrete semiclassical trajectories with the complex initial time, where the imaginary part can be interpreted as quantum tunneling and the real part approximately satisfies the classical trajectories in the three-step model.
The HHG spectrum is then obtained by the Fourier transform of Eq.~(\ref{FTd}). The stationary phase approximation with the saddle point $t=t_r$, which is the return time of the ionized electron, determines the emitted photon energy $\Omega=E_r (t_r)+1.32I_p$ \cite{Lewenstein1994a}, in which $E_r$ is the returning kinetic energy. 
In addition, $d_{gc}(t)$ dominates the Fourier transform over its complex conjugate term, which can be neglected by the rotating wave approximation (RWA).\par
The wave function of the perturbed part $| \Phi_\text{X} (t) \rangle$ satisfies the first-order equation:
\begin{equation} \label{first_order}
 i\frac{\partial}{\partial t}| \Phi_\text{X} (t) \rangle= [\hat{H}_0+\hat{H}_\text{IR}(t)]| \Phi_\text{X}(t) \rangle + \hat{H}_\text{X}(t)| \Phi_\text{IR} (t) \rangle.
\end{equation}
Solving the inhomogeneous differential equation (\ref{first_order}) with the initial condition $| \Phi_\text{X} (t_0) \rangle=0$, we formally get
\begin{equation} \label{perturbed}
| \Phi_\text{X} (t) \rangle=-i\int_{t_0}^{t} dt' \hat{U}_\text{IR} (t,t')\hat{H}_\text{X}(t')|\Phi_\text{IR} (t') \rangle,
\end{equation}
where $\hat{U}_\text{IR} (t,t')$ is the propagator of the system unperturbed by the XUV field according to the Hamiltonian $\hat{H}_0+\hat{H}_\text{IR}(t)$.
We assume that the XUV field only couples the states $|1\rangle$ and $|2\rangle$, and that the dipole transition matrix element $d_{gx}(t)=\langle \Phi_g (t)|z| \Phi_\text{X} (t) \rangle$ can be approximated by stationary phase approximation
\begin{equation}\label{XUVdipole}
 \begin{split}
	d_{gx}(t) \approx&-\int_{t_0}^{t} dt_1 \text{d}_2(p_{(s)}+A(t_1))E_{\text{IR}}(t_1) \\
	& \times\tilde{a}_{\text{pr}}(t,t_1)\tilde{a}_{\text{rec}}(t,t_1) a_\text{xuv}(t,t_1),
 \end{split}
\end{equation}
where the amplitude $a_\text{xuv}$ reads 
\begin{equation} \label{hole_transition}
a_\text{xuv}(t,t_1)=\int_{t_1}^{t}dt_2z_{12}E_\text{XUV}(t_2)e^{ i\Delta \varepsilon t_2}
\end{equation}
with $\Delta \varepsilon=\varepsilon_2-\varepsilon_1$, 
and the modified propagation and recombination amplitudes are 
   \begin{equation} \label{dipole_coefficient_XUV}
   \begin{split}
   \tilde{a}_{\text{pr}}(t,t_{1})&=\bigg(\frac{2\pi i}{t-t_{1}}\bigg)^{1/2}e^{-iS_2(t,t_1,p_{(s)})}\\
   \tilde{a}_{\text{rec}} (t,t_{1})&= \text{d}_1^*(p_{(s)}+A(t)),
   \end{split}
   \end{equation}
where $S_2(t,t_1,p_{(s)})=S_1(t,t_1,p_{(s)})+\Delta \varepsilon t$.
The physical interpretation is that the outer-shell electron is ionized at $t_1$ by the laser field and then propagates in the laser-dressed continuum. At $t_2$, the inner-shell electron is excited to the outer-shell hole, such that the electronic hole is transferred from outer to the inner shell. Finally, the electron in the continuum recombines to the inner-shell hole at time $t$. 
The term $a_\text{xuv}(t,t_1)$ can be interpreted as the transition amplitude of the inner-shell electron to the outer orbital during the excursion of the tunnel ionized electron is in the continuum. 
We will analyze how a given time delay, $\tau$, (or the phase delay, $\delta\equiv 2\pi\tau/ T_\text{IR}$) between the XUV and the IR field affects the HHG spectrum. Therefore, we will write out the time arguments  explicitly $a_\text{xuv}(t,t_1,\tau)$ where we find it necessary. 
It can be shown that the maximal energy of the emitted photons reach $3.17U_p+\Delta\varepsilon+1.32I_p$ in the XUV-assisted HHG processes. \par
To incorporate the XUV driven transition in the strong-field approximation, we redefine the phase factor $S_2$ by moving the XUV transition matrix element into the exponent:
    \begin{equation}
    	S_{2,X}(t,t_1,p_{(s)},\tau)=S_{2}(t,t_1,p_{(s)})+S_{X}(t,t_1,\tau),
    \end{equation}
    where 
    \begin{equation} \label{S_XUV}
    	S_{X}(t,t_1,\tau)=i\ln a_\text{xuv}(t,t_1).
    \end{equation}
Then, the stationary phase requirement with respect to the ionization time $t_1$ yields an additional contribution $\partial S_{X}(t,t_1,\tau)/\partial t_1 =- i \eta(t_1,\tau)$, so that the condition of stationary phase reads
\begin{equation} \label{second_stationary}
\pp{S_{X}}{t_1}\bigg|_{t_{1(s)}}=-\frac{[p+A(t_{1(s)})]^2}{2}-I_p -i \eta (t_{1(s)},\tau) =0,
\end{equation} 
where
\begin{equation}
\eta (t_1,\tau) =\frac{1}{a_\text{xuv}(t,t_1)}\pp{a_\text{xuv}(t,t_1)}{t_1}.
\end{equation} 
If $a_\text{xuv}$ changes slowly so that $a_\text{xuv}(t,t_{1(s)}) \approx a_\text{xuv}(t,t_i)$ and $\eta (t_{1(s)},\tau) \approx \eta (t_{1},\tau) \ll I_p$, the contribution of the XUV fields to the stationary phase equation [Eq.\ (\ref{second_stationary})] can be neglected and $a_\text{xuv}$ can be treated as a constant with respect to $t_1$ in Eq.\ (\ref{XUVdipole}). 
The assumption of the slowly-varying function $a_\text{xuv}$ mentioned above can be expressed as
\begin{equation} \label{slowly_varying_condittion}
\begin{split}
\bigg|\pp{a_\text{xuv}}{t_1}(t,t_1=t_i)\bigg| &\ll \bigg|\frac{1}{\text{Im}(t_{1(s)})} a_\text{xuv}(t,t_i)\bigg| \\
\bigg|\pp{a_\text{xuv}}{t_1}(t,t_1=t_i)\bigg| &\ll |I_p a_\text{xuv}(t,t_i)|,
\end{split}
\end{equation}
which indicates that $a_\text{xuv}$ has to vary slowly on the time scales of the tunneling time and $I_p^{-1}$. For XUV pulses satisfying Eq.\ (\ref{slowly_varying_condittion}), the dipole expectation value Eq.\ (\ref{XUVdipole}) can then be approximated by applying a series of saddle-point approximations, similar to those that led to Eq.\ (\ref{IR_factorization}). In this simplified case we get
\begin{equation} \label{ds}
d_{gx}(t) \approx \sum_{t_i} a_\text{ion} (t,t_i) \tilde{a}_\text{pr}(t,t_i) \tilde{a}_\text{rec}(t,t_i) a_\text{xuv}(t,t_i).
\end{equation} 
The XUV resonant excitation has two major implications for the semiclassical description: the additional phase in $\tilde{a}_\text{pr}$ representing the inner hole propagation and the addition of transition amplitude $a_\text{xuv}$, which is a real quantity in the RWA.
If the pulse shape $E_{X}$ of the XUV field $E_\text{XUV}(t)=E_{X}(t-\tau)\cos [\omega_x (t-\tau)]$ is short, so that Eq.\ (\ref{slowly_varying_condittion}) is not satisfied, a more detailed analysis is required. For example, if the pulse shape is given by a Gaussian function
\begin{equation} \label{E_Gaussian}
	E_\text{X}(t-\tau)= E_x e^{-(t-\tau)^2/\tau_x^2}
\end{equation}
with central frequency $\omega_x=\Delta \varepsilon$, then
\begin{equation}
   a_\text{xuv}(t,t_1) \approx \frac{z_{12} E_x\tau_x e^{-i\omega_x \tau}}{2} \bigg[\frac{\sqrt{\pi}}{2}-\text{erf}\bigg(\frac{t_1-\tau}{\tau_x}\bigg)\bigg],
\end{equation}
where $\text{erf}$ is the error function, and its contributed dipole phase
\begin{equation}
	\text{Re}[S_{X}(t,t_1,\tau)]=-\omega \tau.
\end{equation}
In the above equation, the RWA is applied and only the negative frequency component of the XUV field is considered:
\begin{equation}
 E_{XUV}^{(-)}(t)=\frac{1}{2}E_\text{X}(t-\tau)\exp[-i\omega_x(t-\tau)].
 \end{equation} 
 In the case of attosecond pulses, $\tau_x$ is small compared to the optical period and the error function grows quickly along the imaginary direction $t_{1(s)}$ when ionization time is close to the center of the XUV pulse $\text{Re}(t_1)\approx\tau$:
 \begin{equation}
    \text{erf}\bigg[i\frac{\text{Im}(t_{1(s)})}{\tau_x} \bigg] \propto \exp \bigg[ \bigg(\frac{\text{Im}(t_{1(s)})}{\tau_x}\bigg)^2 \bigg]
 \end{equation}
 Therefore, $a_\text{xuv}$ should be considered in the stationary requirement for ionization $t_1$. The stationary phase conditions are therefore strongly modified when the attosecond pulse overlaps with the IR induced ionization time. The effect of the pulse-shape of the XUV field on the stationary phase behavior and the HHG spectrum will be discussed in the next section.
  \section{Result and Discussion} \label{Result_and_Discussion}
  We adopt our theory to krypton atoms by matching the binding energies in our 1D model of the $3d$ and $4p$ shells, which implies an ionization potential $I_p=14.0$ eV and resonant excitation energy $\Delta \varepsilon=78.5$ eV. In our 1D model system, the radial dipole transition matrix element is 0.354 a.u. for the hole transition, while the real $3d-4p$ transition is  $3.9\times 10^{-2}$ a.u. in krypton ions.   
  The optical laser intensity is set to $10^{14}$ W/cm$^2$ at a wavelength of 1064 nm. The XUV pulse has a Gaussian shape with a central frequency of 67 times IR-frequency, pulse duration 280 as, and a peak intensity of $10^{12}$ W/cm$^2$. 
  We study two cases: (a) few-cycle IR field + single XUV pulse and (b) flat-top IR field + XUV pulse train. \par
      \begin{figure}[ht]
      	\begin{center}
            \subfigure{\includegraphics[width=0.9\columnwidth,trim=1.5cm 0.5cm 1cm 0.0cm]{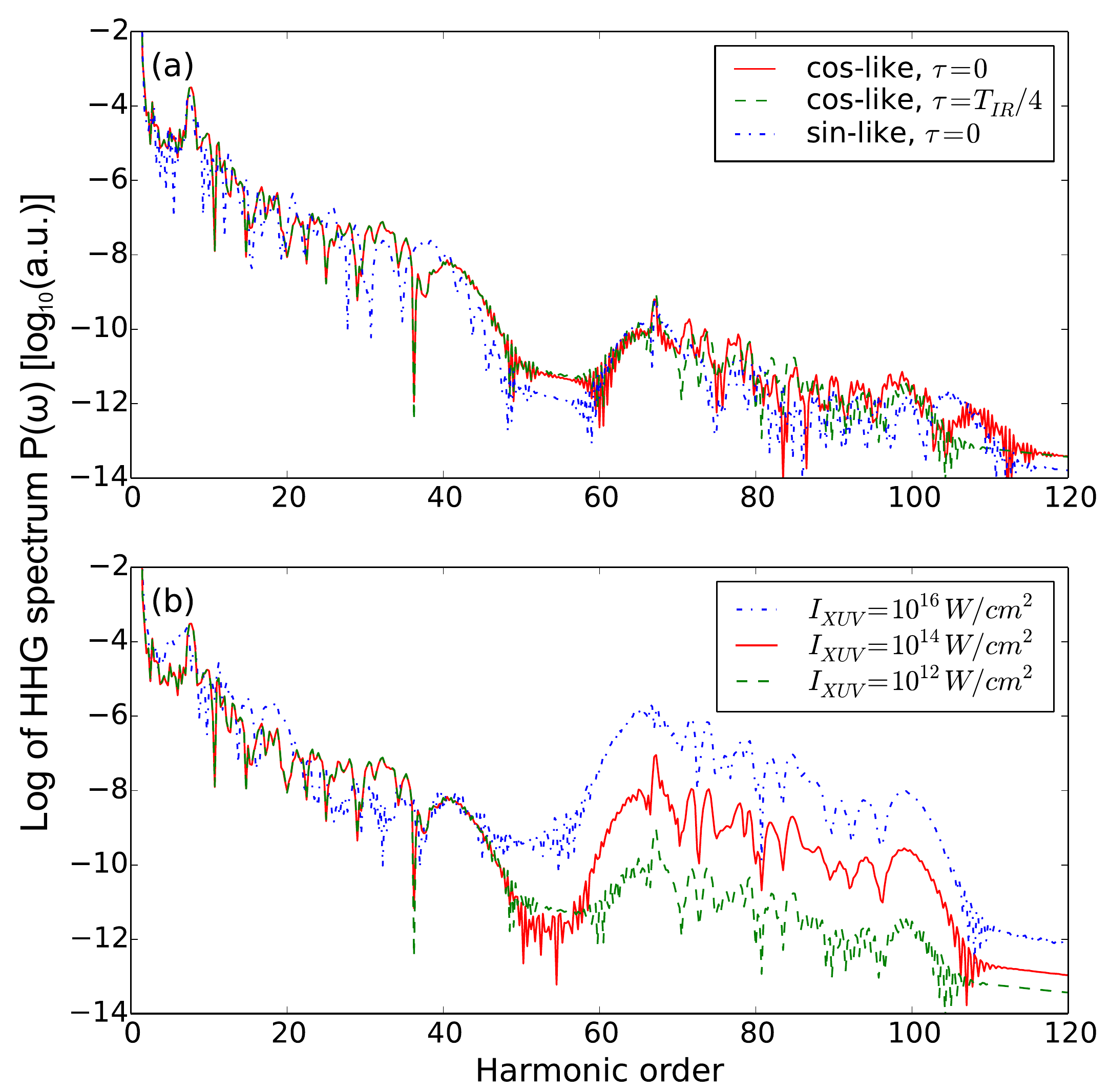}}  
      		\caption{(Color online)
             (a) Comparison of the XUV assisted HHG spectra in logarithmic scale using different four-cycle driving IR fields and a single XUV resonant pulse with different time delay $\tau$: cos-like IR field, $\tau=0$ (red solid curve); cos-like IR field, $\tau=T_{\text{IR}}/4$ (green dashed curve); sin-like IR field, $\tau=0$ (blue dot-dash curve). (b) Comparison of the HHG spectrum using the same cos-like IR field but different XUV intensity at the fixed time delay $\tau=T_{\text{IR}}/4$:
             $I_{\text{XUV}}=10^{12} \text{W}/\text{cm}^2$ (green dashed curve, as the same as the upper panel); $I_{\text{XUV}}=10^{14} \text{W}/\text{cm}^2$ (red solid curve);  $I_{\text{XUV}}=10^{16} \text{W}/\text{cm}^2$ (blue dot-dash curve).}
      		\label{compare_sincos_intensity}
      	\end{center}
      \end{figure}
    \begin{figure*}[ht]
    	\begin{center}
    		\subfigure{\includegraphics[clip,scale=0.60,trim=0cm 0cm 0cm 0cm]{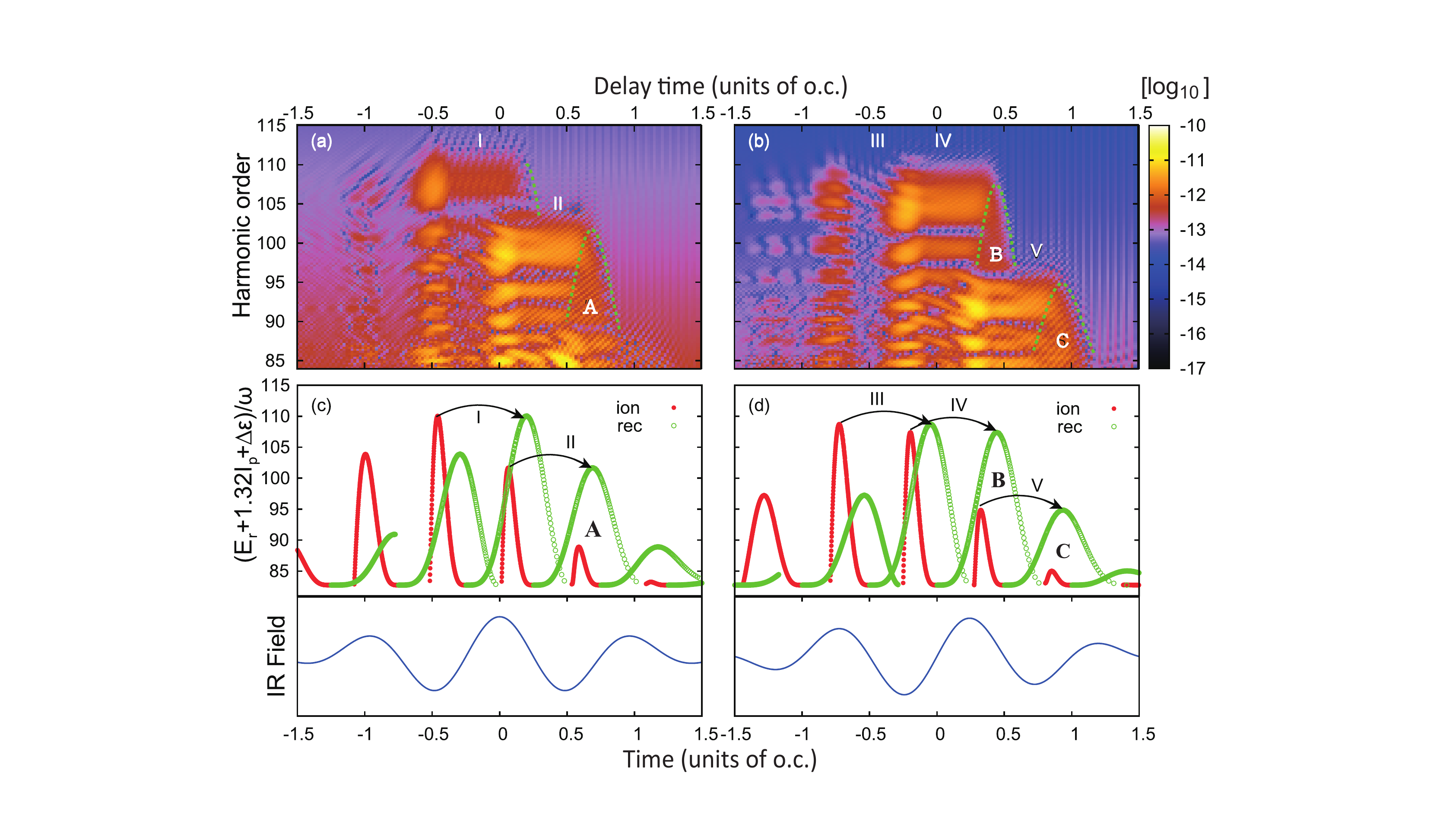}} 
    		\caption{(Color online) The upper panels are power spectra in the second plateau region of HHG  for (a) cos-like and (b) sin-like short IR laser field with single XUV pulse.
    			The lower panels are the return energy (in terms of the harmonic order with recombination to the inner shell) 
    			plotted as a function of the ionization time (red filled circles) and the recombination time (green open circles). 
    			The blue curves in the lower panels are the corresponding IR field: cos-like (left) and sin-like (right). 
    			The arrows in (c) and (d) are the classical trajectories mainly allowed and labeled with \rom{1}-\rom{5}.} 
    		\label{HHG_shortpulse}
    	\end{center}
    \end{figure*}  
    \subsection{SINGLE XUV PULSE + FEW-CYCLE IR FIELD}
  When a few-cycle IR field is applied to an atomic gas, the harmonic peaks in the plateau of the HHG spectrum depend on the carrier envelope phase (CEP) of the IR field. These CEP-dependent structures can be explained as the sum of individual half-cycle bursts and have been observed experimentally \cite{Haworth2006,Ishii2012,Ishii2014}. Here, we consider the HHG spectra with a four-cycle-pulse in addition to a single resonant XUV pulse as shown in Fig.\ \ref{compare_sincos_intensity} (a). The red dashed line shows the HHG spectrum generated by the driving cos-like IR pulse and an XUV pulse with time delay $\tau=0$ defined relative to the peaks of the IR-pulse envelope. The green dashed line shows the result by the same pulses with time delay $\tau= T_{\text{IR}}/4$, where $T_{\text{IR}}$ is the period of the IR field, and the blue dash-dotted line shows the result for a sin-like IR pulse and an XUV pulse with zero time delay, $\tau=0$. In addition to the normal HHG plateau, which runs up to approximately 40 harmonic order and only depends on the CEP of the IR field. In addition, there is a second plateau that originates from the XUV resonant hole transition during the HHG process. The second plateau region is about 3-4 orders of magnitude weaker than the main plateau and could be enhanced by using higher XUV intensity because the transition probability of the core-electronic to the valence vacancy is proportional to the XUV intensity \cite{Buth2011,Buth2012} as shown in Fig.\ \ref{compare_sincos_intensity} (b). The yields of the second plateau can be comparable with the yields of the first plateau with the use of higher XUV intensity such as $10^{16} \text{W}/\text{cm}^2$ [the blue dot-dash line in Fig.\ \ref{compare_sincos_intensity} (b)], but the structure of the first plateau changes because the high-intensity XUV field results in the noticeable depletion of the outer-hole ionic state. In addition to the XUV intensity, the spectrum of the second plateau also depends on the XUV time delay, $\tau$ and the electron trajectories of the HHG process. In Figs.\ \ref{HHG_shortpulse} (a) and \ref{HHG_shortpulse} (b), we show  the 2D plot of HHG spectra in the second plateau region with cos-like and sin-like four-cycle-pulse as a function of time delay, $\tau$, in units of optical cycle (o.c.). Several plateaus are observed with horizontal-stripe structures.\par

For cos-like IR pulse case in Fig.\ \ref{HHG_shortpulse} (a), the main second plateau (\rom{1}) extends up to $110$ harmonic for XUV time delays between $-0.5T_\text{IR}$ and $0.2T_\text{IR}$, while the region (\rom{2}) extends up to $100$ harmonic for XUV time delays between $0$ and $0.7 T_\text{IR}$.
For sin-like IR pulses in Fig.\ \ref{HHG_shortpulse} (b), the second plateau (\rom{4}) extends up to $110$ harmonic for XUV time delays between $-0.3T_\text{IR}$ and $0.5T_\text{IR}$, while the second plateau form a cut-off of harmonic 95 for delay times between $0.5T_{\text{IR}}$ to $0.9T_{\text{IR}}$ (region \rom{5}).
These structures can be understood analzying the classical trajectories. The second plateau can only be generated when the excursion of the valence electron overlaps with the XUV pulse. To explain the structures we analyze the recombination energy as a function of ionization and recombination times that result from the unperturbed case of an acting IR field only. Figures \ref{HHG_shortpulse} (c) and \ref{HHG_shortpulse} (d) plot the recombination energy as a function of a given ionization time by red dots. A classical trajectory of a given ionization time has also a well-defined recombination time, i.e., there is a one-to-one mapping of ionization and recombination times. The recombination energy as a function of recombination time is plotted by the green dots in Figs.\ \ref{HHG_shortpulse} (c) and \ref{HHG_shortpulse} (d). As an example, for the one-to-one mapping we show the corresponding ionization and recombination times for the cut-off trajectories of region \rom{1} of the highest possible energies, which separate the trajectories into two sets of groups: short trajectories, in which the electrons are ionized later and recombine earlier; long trajectories, in which the electrons are ionized earlier and recombine later. If the time delay of the XUV pulse falls in between those times, the plateau region \rom{1} in Fig. \ref{HHG_shortpulse} (a) is generated. Likewise the cut-off trajectories of region \rom{2} are marked in Fig.\ \ref{HHG_shortpulse} (c). A similar analysis holds for the sin-like IR pulse shown in Figs.\ \ref{HHG_shortpulse} (b) and \ref{HHG_shortpulse} (d).
Therefore, the cutoff energy and the peak structures of the second plateau reveal directly the excursion time of the electron trajectories in the strong laser field. 
Clearly, the plateau structure depends critically on the temporal shape of the laser field and one can imagine that the XUV-assisted HHG spectrum can be used to probe the CEP of a few-cycle pulse without the emission of photo- or Auger- electrons \cite{Averbukh2010}.\par
\begin{figure}[ht]
	\begin{center}
        \includegraphics[width=1.2\columnwidth,trim=6.3cm 1cm 1.6cm 0.1cm]{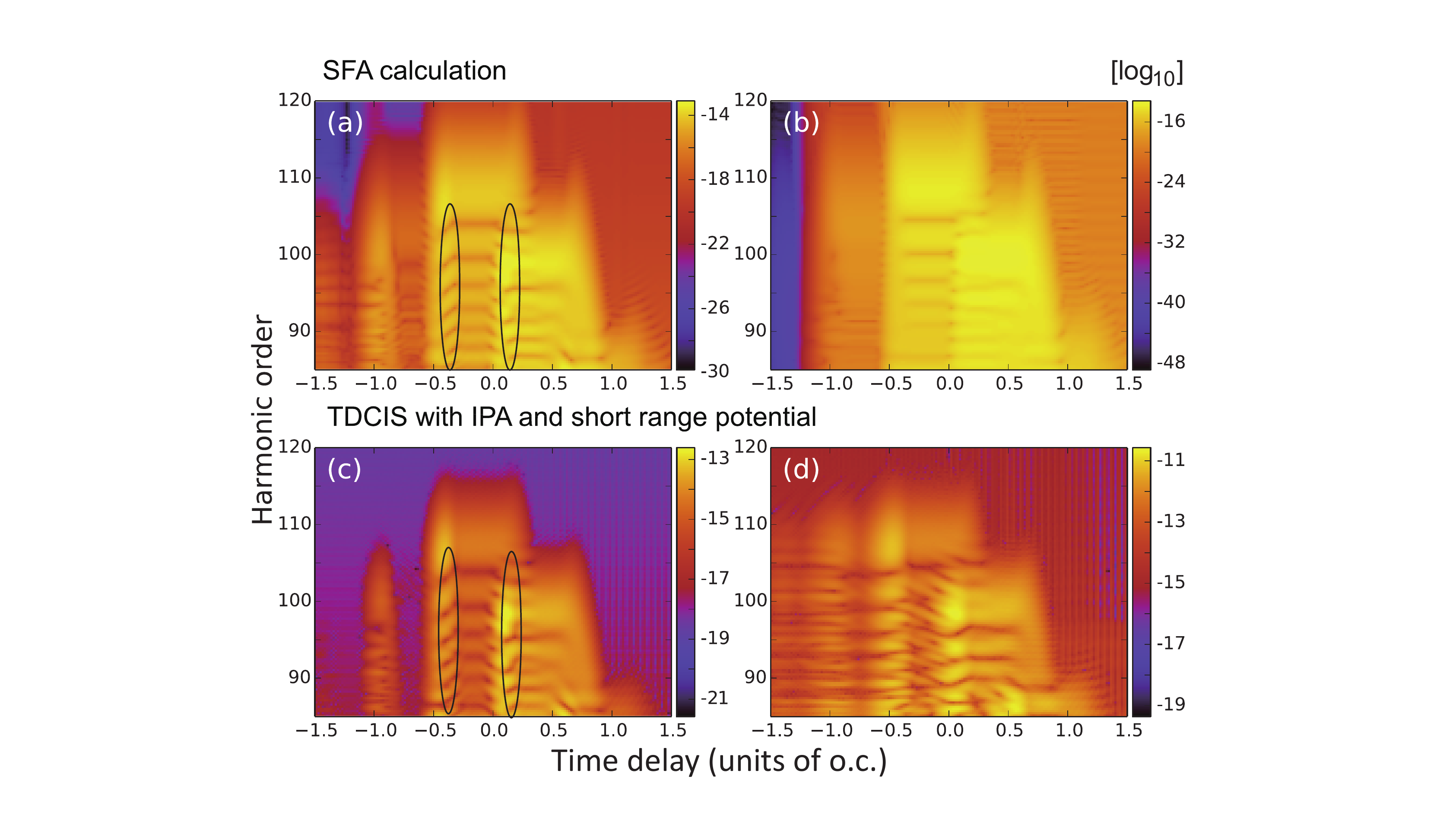} 
		\caption{(Color online) The 2D plot of HHG spectra in the second plateau region for the cos-like four-cycle IR laser field with single XUV pulse as a function of time delay: (a) SFA with SPA for $p$ using Eq.\ (\ref{XUVdipole}); (b) SFA with SPA for both $p$ and $t_1$ using slowly-varying approximation in $a_{xuv}$; (c) TDCIS in the independent particle approximation with the use of short-range potential supporting only two bound states; (d) TDCIS in the independent particle approximation with the use of short-range potential supporting more than two bound states.} 
		\label{HHG_2Dcompare}
	\end{center}
\end{figure}
\begin{figure}[ht]
	\begin{center}
		\includegraphics[width=1.0\columnwidth,trim=6cm 3cm 7cm 4cm]{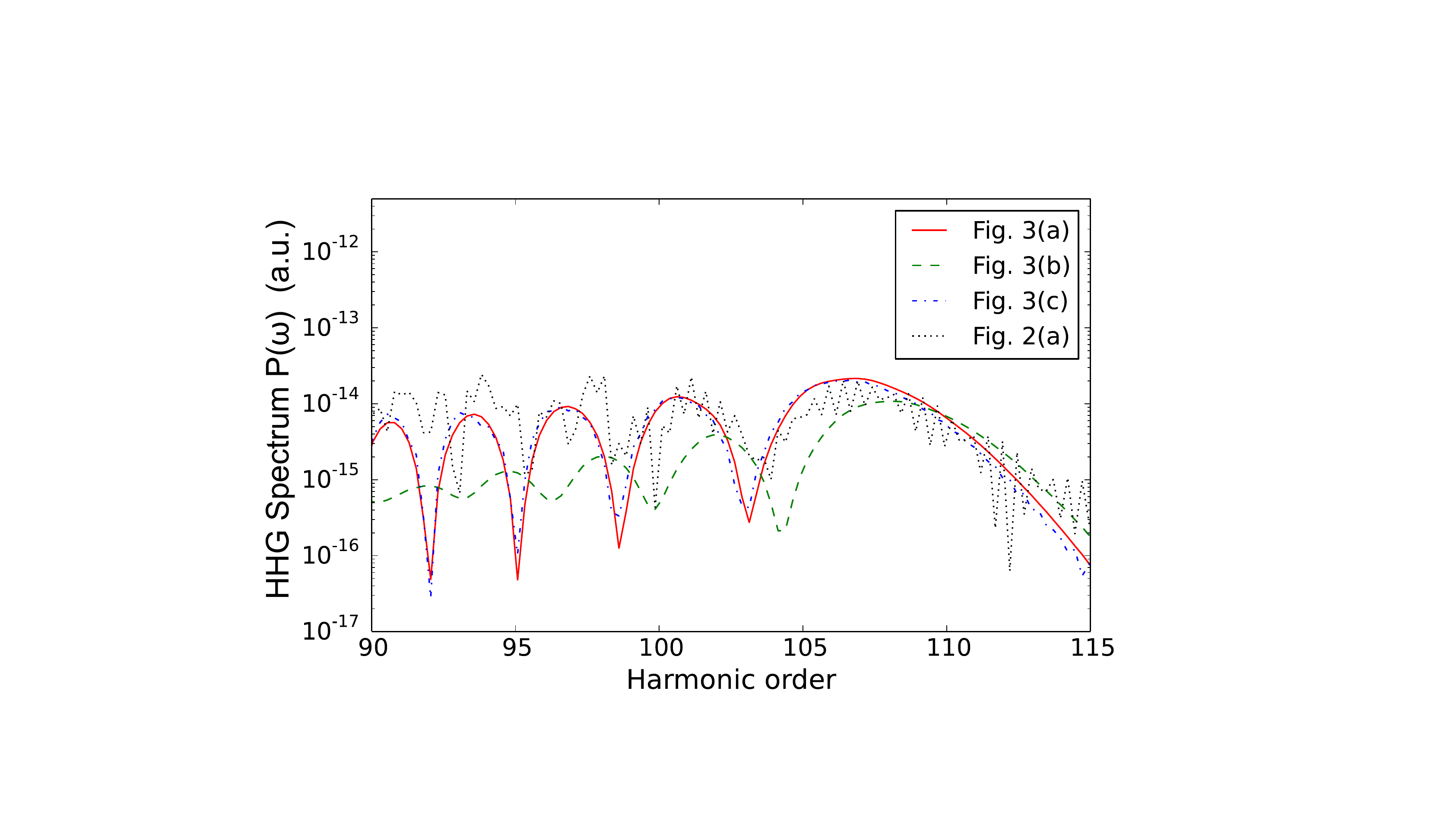} 
		\caption{(Color online) The comparison of HHG spectrum at $\tau= -0.4 T_{\text{IR}}$ for Figs.\ \ref{HHG_2Dcompare} (a)-\ref{HHG_2Dcompare} (c) and Fig.\ \ref{HHG_shortpulse} (a). } 
		\label{same_delay}
	\end{center}
\end{figure}
    To understand the structures of Figs.\ \ref{HHG_shortpulse} (a) and \ref{HHG_shortpulse} (b) in more detail, we perform the semiclassical SFA treatment for the cos-like IR field case. The dipole moment calculation with approximation according to Eqs.\ (\ref{XUVdipole}) and (\ref{hole_transition}) for $a_{xuv}$ leads to Fig.\ \ref{HHG_2Dcompare} (a). The detailed substructures inside the plateaus differ between the TDCIS calculation and the SFA calculation.
    In the case of SFA with Eq.\ (\ref{XUVdipole}) shown in Fig.\ \ref{HHG_2Dcompare} (a), the local minima of the second plateau show positive slopes as a function of time delay for delay times around $\tau=-0.4 T_{\text{IR}}$ and $0.1 T_{\text{IR}}$, as indicated by the regions marked with the black ellipse. In the TDCIS calculation, shown in Fig.\ \ref{HHG_shortpulse} (a), there are more complex substructures. 
    To clarify the origin of the discrepancy between TDCIS and SFA, we simplify our TDCIS to an independent particle approximation (IPA) and use short-range potentials to more closely adapt to the assumptions of SFA. Here we use two kinds of short range potential:
    $V(x)=Ze^{-ax^2}$ with $Z=5.38$ and $a=2.03$ resulting in only two bound states that are occupied by the two electrons [Fig.\ \ref{HHG_2Dcompare} (c)]; $V(x)=Ze^{-ax^2}/\sqrt{x^2+b^2}$ with $Z=1.07$, $a=0.01$ and $b=0.702$ supporting a larger number of bound excitations [Fig.\ \ref{HHG_2Dcompare} (d)]. In the IPA with two bound states only, the pattern within the region marked by the black ellipse in the Fig.\ \ref{HHG_2Dcompare} (c) is similar to the one of Fig.\ \ref{HHG_2Dcompare} (a). For example, we compare the HHG spectrum at $\tau=-0.4 T_{\text{IR}}$ as shown in Fig.\ \ref{same_delay}. Because the asymptotic behavior rather than the analytical from of the of the bound state wavefunction is known (see the appendix), the HHG spectrum in SFA model shows relative photon yields up to a normalization factor. Here, the data of the SFA model from Fig.\ \ref{HHG_2Dcompare} (a) (blue dash-dot curve) is normalized to the scale of the data of the IPA with potential supporting only two bound states from Fig.\ \ref{HHG_2Dcompare} (c) (red solid curve), and these two curves fit very well with the same peak structure. On the contrary, when a potential supporting more than two bound states is used, a more complex interference structure in the plateaus is visible, as shown in Fig.\ \ref{HHG_2Dcompare} (d). Thus, we argue that the complex interferences in the lower energy part of the plateaus in Fig.\ \ref{HHG_shortpulse} (a) are due to bound excited state population, while the more energetic part of the plateaus are well described by the SFA model and exhibit (i) clear horizontal stripes and (ii) a positive slope on the left side of the plateaus shown in Figs.\ \ref{HHG_2Dcompare} (a) and \ref{HHG_2Dcompare} (c). 
Indeed, the HHG spectrum at $\tau=-0.4 T_{\text{IR}}$ obtained from Fig.\ \ref{HHG_shortpulse} (a) shown as the black dot curve in Fig.\ \ref{same_delay} presents complex oscillation along the line representing the SFA model and the IPA with potential supporting only two bound states. We have found that the detailed structures in the plateaus are independent of depletion of the ground state by direct ionization with the XUV field by setting the XUV field equal to zero in Eq. (\ref{TDCIS_ground}). Moreover, the fine structure in the plateaus is not affected by turning off the XUV interaction in term $\circled{3}$ of Eq.\ (\ref{TDCIS_nodecay}) so that we can conclude that the structure is not resulting by interchannel coupling mediated by the XUV field. 
Therefore, any bound excited state population is a result of strong field excitation by the IR field rather than the coupling with the XUV field, at least in the considered XUV intensities. \par
If $a_{xuv}$ is assumed to be slowly varying, i.e. if we adopt the approximation for the dipole moment of Eq.\ (\ref{ds}), then the stripes at delay times of around $\tau= -0.45 T_{\text{IR}}$ and $-0.35 T_{\text{IR}}$ with positive slopes vanish, as shown in Fig.\ \ref{HHG_2Dcompare} (b). There is a huge range of orders of magnitude in Fig.\ \ref{HHG_2Dcompare} (b) because the lack of classical trajectories in certain areas of the plot makes the background signal weaker. In the region where the positive slopes vanish such as $\tau= -0.4 T_{\text{IR}}$, the HHG spectrum plotted as a green dash curve in Fig.\ \ref{same_delay} gives different peak structures compared with the SFA without the slowly varying approximation (blue dash-dot curve).
These time delays $\tau$ are close to the ionization time $t_i$ of the classical trajectories. 
Clearly, to fully explain the structure of positive slopes structure of positive slopes marked by the black ellipses in Figs.\ \ref{HHG_2Dcompare} (a) and \ref{HHG_2Dcompare} (c), the phase contribution $S_{X}(t,t_1,\tau)$ has to be taken into consideration within the XUV field in the stationary phase approximation. 
    The solution of the stationary phase equation including $S_X$ yields 
    \begin{equation}
    t_{1(s)}=t_{i}-\frac{i\eta_0^2\pm\sqrt{-\eta_0^4-2[|E(t_i)|^2+i2\eta_0^3][I_p+i\eta_0]}}{|E(t_i)|^2+i2\eta_0^3},
    \end{equation}
    where $\eta_0=\eta(\tau,\tau)$. The values of the solution determined by the laser parameters we use are $t_{1(s)}-t_{i}=1.91 + 15.2i$ and $4.72-19.8i$, and the one in the upper complex plane should be chosen so that the integration can turn out a simple Gaussian integral along a suitable contour, which is parallel to the real axis. Compared with the original complex ionization time $t_{1(s)}=t_{i}+\sqrt{2I_p}/|E(t_i)|$, the new term $S_X$ changes both the real part and imaginary part of the complex ionization time. Because $\text{Re}(t_{1(s)})$ does not deviate from $t_i$ much, we can assume $\text{Re}(t_{1(s)})=t_i$ so that the time integration $a_{xuv}(t_r,t_{1(s)},\tau)$ inside $S_X(t_r,t_{1(s)},\tau)$ can be split into two parts: $a_{xuv}(t_r,t_i,\tau)$ with $\text{Arg}[a_{xuv}(t_r,t_i,\tau)]=i\omega_x \tau$, corresponding the resonant hole transfer during the outer electron in the continuum; and $a_{xuv}(t_i,t_i+i\text{Im}(t_{1(s)}),\tau)$, the resonant excitation during the tunneling process. With the Gaussian shaped XUV pulse of Eq.\ (\ref{E_Gaussian}), this complex transition amplitude can be approximated by $a_{xuv}(t_r,t_{1(s)},\tau)\approx \exp (i\omega_x\tau) (R+iI)$, where
        \begin{equation}
        R=\frac{\tau_x \sqrt{\pi}}{2}+ \exp\bigg[ \frac{(\text{Im}(t_{1(s)}))^2}{\tau_x^2} \bigg] (\tau-t_i)
        \end{equation}
        and
        \begin{equation}
        I=-\int_{0}^{\text{Im}(t_{1(s)})} dt_1' \exp \bigg(\frac{{t'}_1^2}{\tau_x^2}\bigg).
        \end{equation}
    The additional phase contributed from the resonant excitation during the tunneling process is equal to 
    \begin{equation}
      \text{Re}[S_{X}(t_r,t_{1(s)},\tau)]-\text{Re}[S_{X}(t_r,t_i,\tau)] = -\pi/2+ \arctan(R/I).
    \end{equation}
    In the case $\tau=t_i$, $R\approx 10.4$ is smaller than the absolute value of $I$, which is about $30.5$, so we obtain the approximation 
    \begin{equation} \label{RE_S}
       \text{Re}(S_{X}(t_r,t_{1(s)},\tau))+\omega_x\tau \approx -\frac{\pi}{2}+\frac{R}{I}=-A-B(\tau-t_i),
    \end{equation}
    where
    \begin{equation}
    A=\frac{\pi}{2} + \frac{\tau_x \sqrt{\pi}}{2\int_{0}^{\text{Im}(t_{1(s)})} dt_1' \exp \big(\frac{{t'}_1^2}{\tau_x^2}\big)}
    \end{equation}
    and
    \begin{equation}
     B=\frac{\exp\big[ \frac{(\text{Im}(t_{1(s)}))^2}{\tau_x^2} \big] }{\int_{0}^{\text{Im}(t_{1(s)})} dt_1' \exp \big(\frac{{t'}_1^2}{\tau_x^2}\big)}.
    \end{equation}
    Therefore, when the XUV pulse is close to the start of the classical trajectory, there is an additional phase which is linear to the time delay $\tau$, as accessed in Eq.~(\ref{RE_S}). 
    When the XUV pulse is applied far from the IR tunnel ionization peaks, $\eta(t_1,\tau)$ is close to zero so that the stationary phase behavior is independent of $\tau$.\par
\begin{figure}[ht]
	\begin{center}
		\includegraphics[width=1.0\columnwidth,trim=0.2cm 0cm 0cm 0cm]{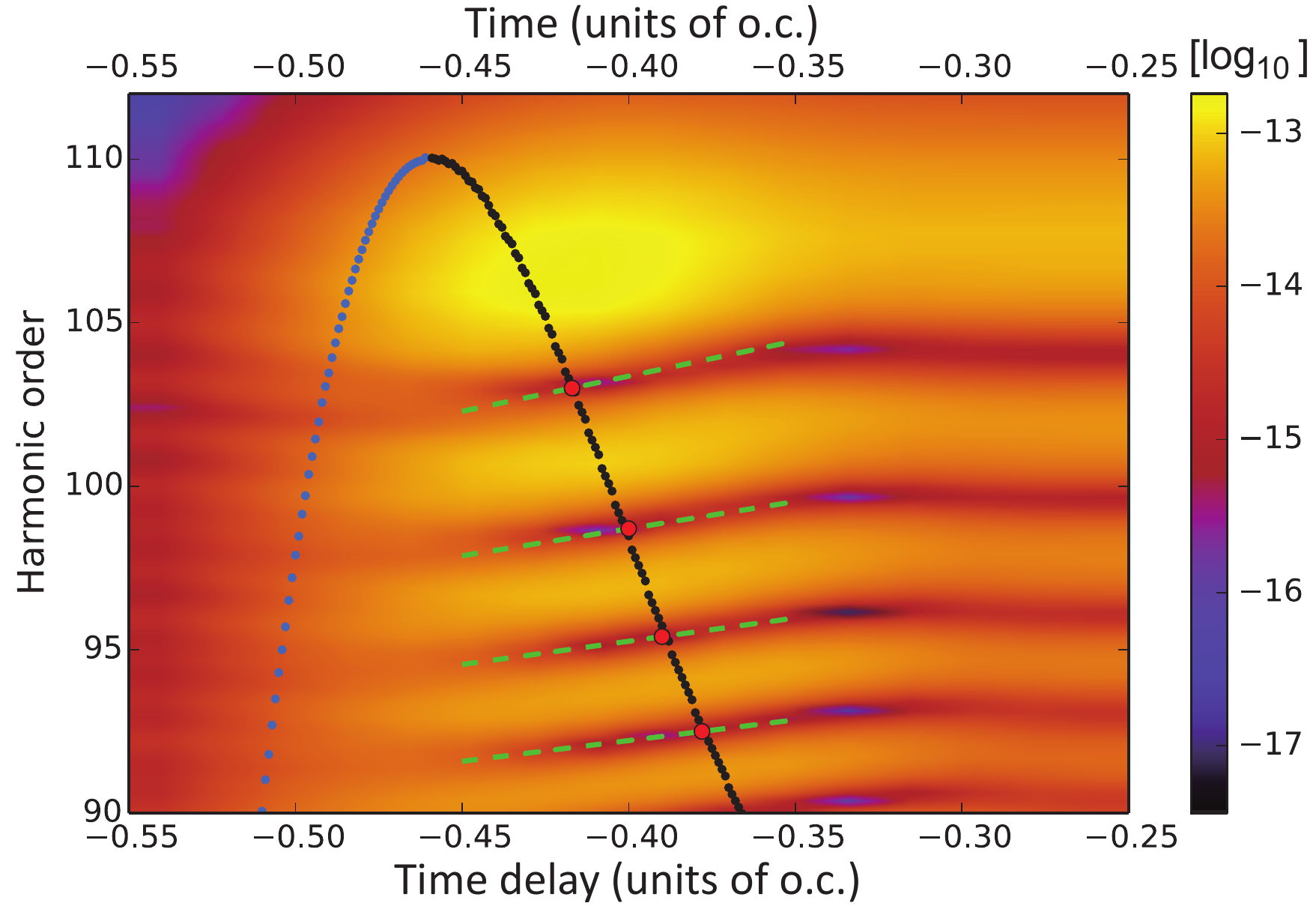} 
		\caption{(Color online) Enlarged region of Fig.\ \ref{HHG_2Dcompare} (a) for delay times of roughly -0.4 optical cycles. In this region, the stripe structure shows a positive slope. The dotted curve shows emitted photon energy in the second plateau as a function of ionization time. Long trajectories, with ionization times earlier than the cut-off trajectory (trajectory with maximal return energy) marked in blue, and short trajectories marked in black. The green dashed lines mark the minima of the interference pattern predicted by Eq.\ (\ref{interference_pattern}) for $\tau \approx t_i$. } 
		\label{stripe_distortion}
	\end{center}
\end{figure}

 
    The interference pattern in the second plateau originates from the interference between short and long trajectories. In this case the additional $\tau$-dependent phase in Eq.\ (\ref{RE_S}) can be resolved as shown in Fig.\ \ref{stripe_distortion} showing a zoomed-in region of Fig.\ \ref{HHG_2Dcompare} (a). The pointed curves in Fig.\ \ref{stripe_distortion} indicate the emitted photon energy as a function of ionization time for long (blue dotted line) and short (black line) trajectories. When $\tau$ is close to $t_{i}^s$ the ionization time of the short trajectories, the phase of the HHG interference pattern is
    \begin{equation} \label{interference_pattern}
    \begin{split}
    &-\Omega (t_{r}^l-t_{r}^s) +S_2 (t_{r}^l,t_{i}^l,p_{(s)})\\&-S_2 (t_{r}^s,t_{i}^s,p_{(s)}) +A^s + B^s (\tau-t_{i}^s),
    \end{split}
    \end{equation} 
    where the superscripts $l$ and $s$ label the long and short trajectories, respectively. Here the resonant excitation does not happen for the long trajectory due to the nonoverlapping between the XUV field and $t_{i}^l$, so
    \begin{equation}
       \text{Re}(S_{X}(t^l_r,t^l_{1(s)},\tau))=\text{Re}(S_{X}(t^l_r,t^l_{i},\tau))=-\omega_x\tau.
    \end{equation} 
    Therefore, the valleys of the HHG signal show a linear drift with positive slopes $B^s/(t_{r}^l-t_{r}^s)$, as indicated by green dashed lines, which fits well with the local minima of the data shown in red color and cross the black dot line. When $\tau$ is close to $t_{i}^l$, the interference pattern disappears because the XUV pulse is applied at a time in the optical cycle, that does not support the production of the short trajectory.\par
    
    It is worth noting that if a cw XUV pulse is used, the XUV resonant excitation mainly happens during the electron excursion. In this case the slowly varying XUV amplitude approximation is valid. However, if the XUV pulse is short and if it overlaps with the start of an electron trajectory, the effective tunneling time may change and the XUV-assisted HHG is enhanced [c.f. Figs.\ \ref{HHG_shortpulse} (a) and \ref{HHG_shortpulse} (b)], where the plateaus are brighter on the left side, such as $\tau=-0.5 \pi$ and $0$ in Fig.\ \ref{HHG_shortpulse} (a). This effect and the positive slopes of the interference between short and long trajectories demonstrate the {\it invalidity of the slowly-varying approximation} of $a_{\text{xuv}}$ and it presents a complication to probe the exact tunneling and return times of electrons \textit{in situ} in a strong laser field by a single attosecond pulse.  
The situation is quite similar to other cases where classical pictures are used as a starting point for perturbative expansions on top of the SFA model. For example, the one-color SFA was perturbed by a weak second harmonics field for the purpose of \textit{in situ} characterization of attosecond pulses. If one assumes classical trajectories, the second harmonic contributes only a phase correction in the action \cite{Dudovich2006}. However, using more refined quasiclassical trajectories, which includes the imaginary part of the saddle points, shows that also the tunneling processes is modified by the second harmonic field and that the behavior of the high-order even harmonics is not well suited for pulse characterization \cite{Dahlstrom2011}. 
  \subsection{XUV PULSE TRAIN + CW IR FIELD}
    \begin{figure}[ht]
    	\begin{center}
    		\includegraphics[width=1.0\columnwidth,trim=5cm 5cm 5cm 7cm]{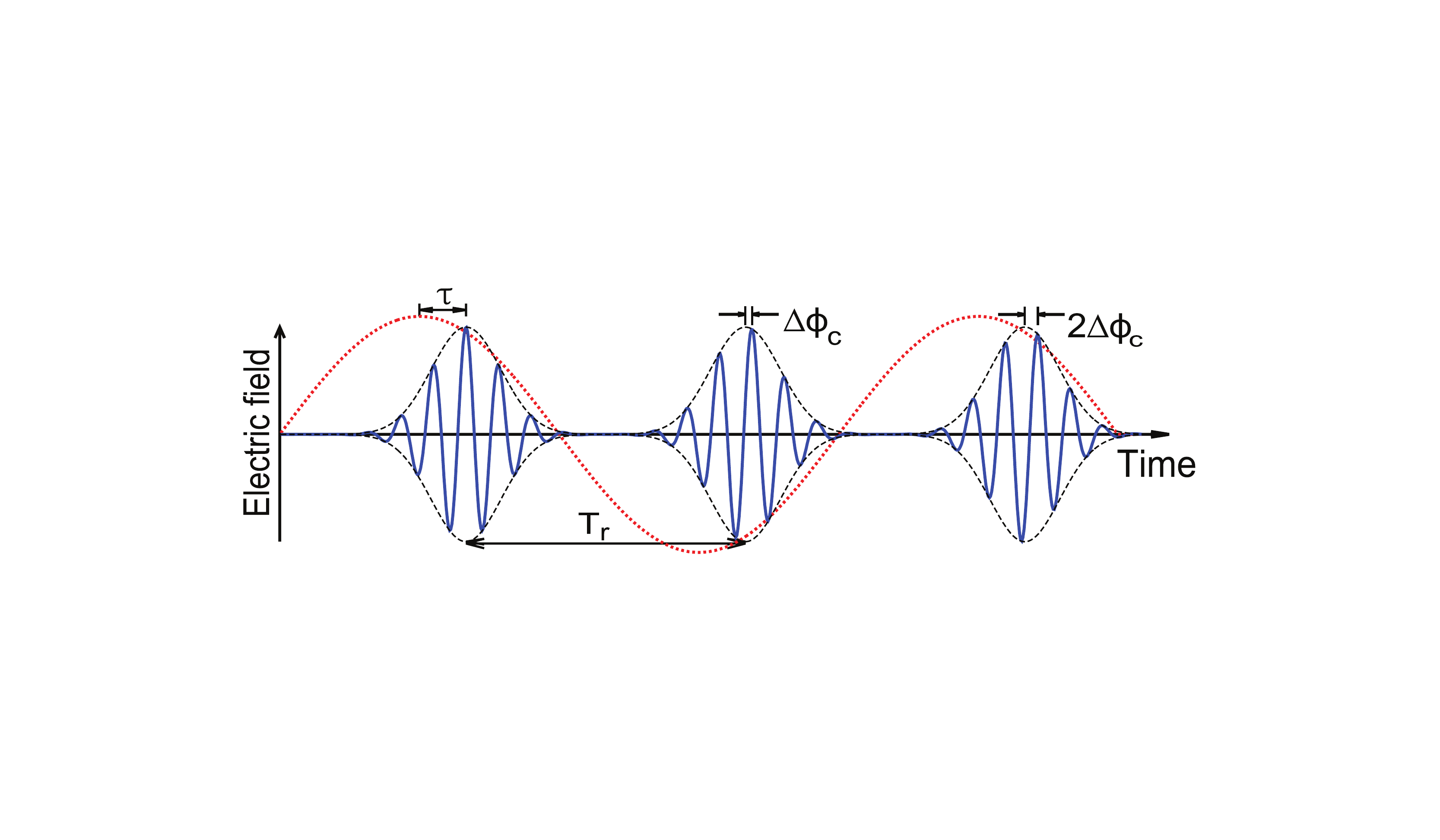}  
    		\caption{(Color online) Electric field of the IR field (red dash-dot line) and the XUV pulse train (blue solid line). 
    			If the XUV pulse train is generated via HHG in atomic gases, then $\Delta \phi_c=\pi$ and the repetition period is equal to half cycles of the IR field, $T_{r}=T_\text{IR}/2$.} 
    		\label{train_schematic}
    	\end{center}
    \end{figure}
    \begin{figure}[ht]
    	\begin{center}
    		\includegraphics[width=1.0\columnwidth,trim=0cm 0cm 0cm 0cm]{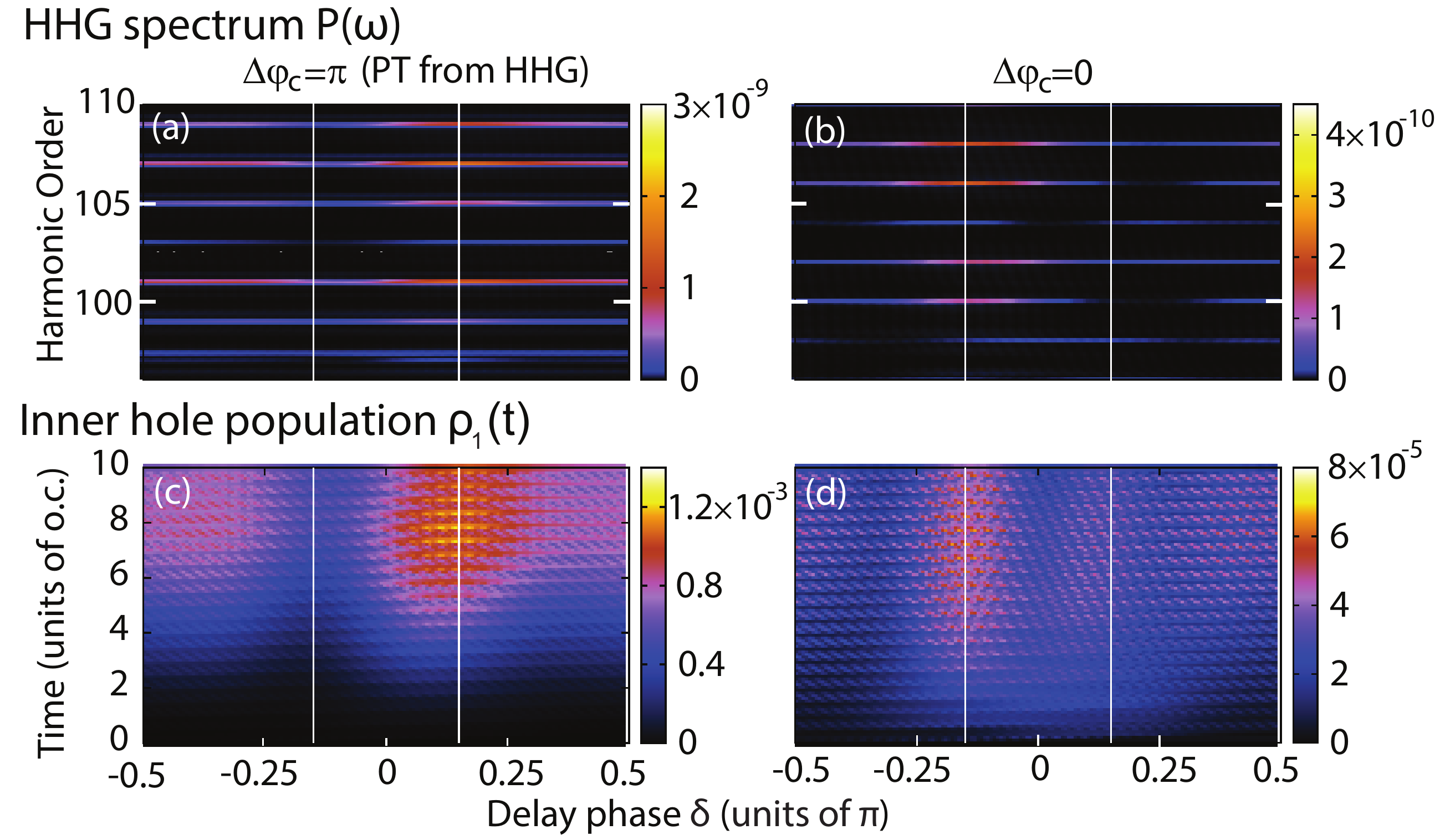}
    		\caption{(Color online) The HHG in the second plateau region with two different kinds of XUV pulse train, $\Delta \phi_c$ is $\pi$ (left) and $0$ (right), as a function of delay phase $\delta$.
    			(a) and (b) HHG power spectra in the second plateau region with acceleration form. (c) and (d) Inner-shell population $\rho_{1}(t)$ as the function of delay phase $\delta$.
    			The white lines on $\delta=-0.15\pi$ and $\delta=0.15\pi$ indicate the local maximum and minimal area of the HHG spectrum.} 
    		\label{HHG_timedelay_phase}
    	\end{center}
    \end{figure}
    \begin{figure}[ht]
    	\begin{center}
    		\includegraphics[clip,width=1.0\columnwidth,trim=0cm 0cm 0cm 0cm]{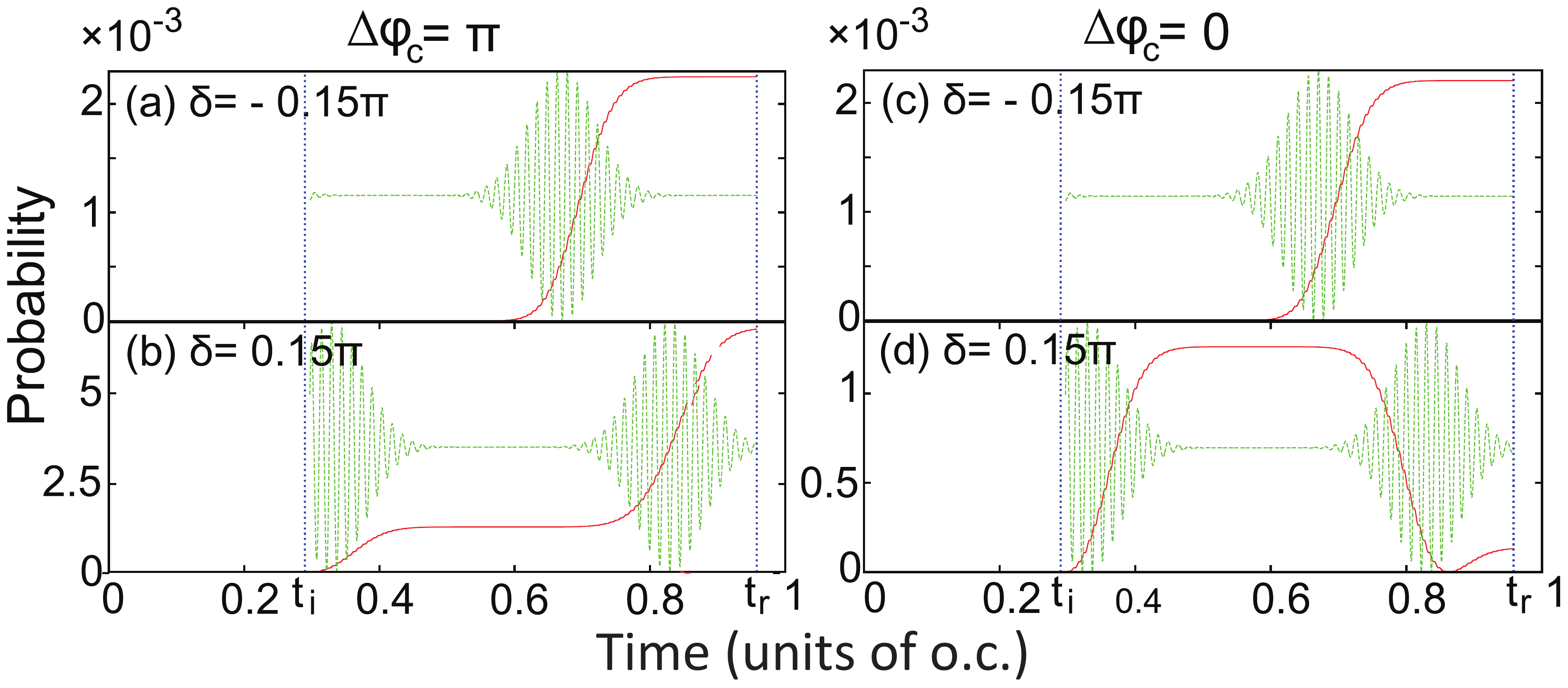}  
    		\caption{(Color online) $|a_\text{xuv}(t,t_i,\tau)|^2$ curve (red lines) from $t=t_i$ to $t_r$ (blue dash line) for XUV pulse trains (green curves) with $\Delta \phi_c=\pi$ (left) and $\Delta \phi_c=0$ (right).
    			Here $\delta$ is chosen $-0.15\pi$ and $0.15\pi$ (white lines in Fig.\ \ref{HHG_timedelay_phase}).
    			When $\delta=-0.15\pi$, the parent ion interacts with the XUV-PT once and both of these two cases show the same curve in (a) and (c).
    			When $\delta=0.15\pi$, the parent ion interacts with the XUV-PT twice and different CEO values show different features on the curves in (b) and (d).} 
    		\label{time_population_phase}
    	\end{center}
    \end{figure}
  When an XUV pulse train (PT) is applied in addition to a driving IR field for the HHG, the periodic interaction with the XUV pulse can result in a coherent accumulation of the inner-shell hole population. 
  As common in typical experimental set ups, we suppose that the individual bursts of the XUV-PT are separated by half the IR laser period (see Fig.\ \ref{train_schematic}), with a relative carrier-envelope offset (CEO) of $\Delta \phi_c =\pi$, so that the electric field of the XUV PT reads
  \begin{equation}
  E_{\text{XUV}}(t)=\sum_{m} E_{\text{X}} (t-\tau_m) \cos [\omega_{x} (t-\tau_m)-m\Delta \phi_c],
  \end{equation}
  where $\tau_m=\tau+mT_{\text{IR}}/2$ and $m$ is integer.
  The resulting second plateau of the HHG spectrum is shown in Fig.\ \ref{HHG_timedelay_phase} (a).
  Here on the $x$-axis, we define the phase delay $\delta \equiv 2\pi\tau/T_\text{IR}$, which is directly linked to the time delay of the IR peak intensity to the envelope of the as pulse, in multiples of $\pi$. Like the first plateau, the second plateau also contains only odd harmonic orders. There is $\pi$ phase shift of the same pathway with the opposite kinetic momentum in adjacent half cycles due to sign-change of the monochromatic driving IR field,
  \begin{equation}
  E_{\text{IR}} (t+\frac{T_{\text{IR}}}{2} )= - E_{\text{IR}} (t).
  \end{equation}
  Here, both the dipole transition with the pulse train,
  \begin{equation}  \label{XUV_symmetry}
  z_{12}E_{\text{XUV}}^{(-)} (t+\frac{T_{\text{IR}}}{2})= z_{12}E_{\text{XUV}}^{(-)} (t) e^{i\Delta \phi_c},
  \end{equation}
  and the opposite parity between the inner and the outer shell,
  \begin{equation}
  d_1^{*}(-k')d_2(-k)=-d_1^{*}(k')d_2(k),
  \end{equation}
  induce additional $\pi$ phase shifts in the pathway of the second plateau in the next half cycles. Therefore, the net phase difference between the pathway from two adjacent half cycle is $\pi$, which results in destructive interference in the even harmonics. 
  The HHG spectrum of the second plateau however exhibits a phase-delay dependence: When the peak of the XUV pulse is in the second and fourth quarter cycle ($\delta>0$), the spectrum shows higher intensity as does the inner hole population $\rho_1 (t)$ shown in Fig.\ \ref{HHG_timedelay_phase} (c) also shows higher population in the same region.
  This can be understood by analyzing the hole transition probability $|a_\text{xuv}(t,t_i,\tau)|^2$, from the outer to the inner shell for each trajectory for $t \in [t_i,t_r]$. 
  As the example study the case for $\delta=0.15\pi$ and $-0.15\pi$, indicated as white lines in Fig.\ \ref{HHG_timedelay_phase}, and the cutoff trajectory, $[t_{i,\text{cutoff}}=0.29T_\text{IR},t_{r,\text{cutoff}}=0.958T_\text{IR}]$.
  The classical trajectory with an analysis presented in Fig.\ \ref{time_population_phase} (a) and (b), showing the transition probability $|a_\text{xuv}(t,t_i,\tau)|^2$ and the XUV field for these two phase delays.
  For $\delta=0.15\pi$ (Fig.\ \ref{time_population_phase} (a)), there is only one XUV pulse in the time interval $[t_{i,\text{cutoff}},t_{r,\text{cutoff}}]$ corresponding to the excursion time of the electron in the continuum for the cut-off trajectory, and the transition probability increases during the XUV field as shown and $|a_\text{xuv}(t_{r,\text{cutoff}},t_{i,\text{cutoff}},\tau)|^2 \approx \gamma$, where 
  \begin{equation} 
  	\gamma = \frac{1}{4} \bigg|z_{12}\int_{-\infty}^{\infty} dt'E_{\text{X}} (t'-\tau)\bigg|^2.
  \end{equation}
  For $\delta=-0.15\pi$, there are two XUV pulses in the time interval between ionization and recombination of the cut-off trajectory $[t_{i,\text{cutoff}},t_{r,\text{cutoff}}]$. 
  The final transition probability is the coherent sum of the transition amplitudes of both contributions and can be written as
  \begin{align}
  |a_\text{xuv}(t_{r,\text{cutoff}},t_{i,\text{cutoff}},\tau)|^2 &\approx \gamma |1+e^{i(\Delta \phi_c + \Delta \varepsilon T_{\text{IR}}/2)}|^2  \nonumber \\
  &=\gamma |1+e^{i(\Delta \phi_c + \pi)}|^2. \label{axuv_phase}
  \end{align}  
  Here $\pi$ in the second line corresponds to the free propagation of the holes, and can be canceled by $\Delta\phi_c=\pi$. 
  Therefore, the transition amplitudes resulting from two consecutive XUV pulses are constructively added so that the time dependent transition probability is doubled as compared to the case for $\delta=0.15\pi$.
  For $\Delta \phi_c=0$, the $\pi$ phase difference due to the free propagation can not be canceled: 
  For $\delta=0.15\pi$, there is only one XUV pulse in the time interval $[t_{i,\text{cutoff}},t_{r,\text{cutoff}}]$ as shown in Fig.\ \ref{time_population_phase} (c), so the behavior of the transition probability is the same as for $\Delta \phi_c=\pi$.
  For $\delta=-0.15\pi$, the transition probability increases during the first XUV pulse and then decreases during the second XUV pulse as shown in Fig.\ \ref{time_population_phase} (d). Therefore, this destructive accumulation reflects lower HHG yields for $\delta=-0.15\pi$ than at $\delta=0.15\pi$ as seen in Fig.\ \ref{HHG_timedelay_phase} (b) and (d).
  Moreover, the second plateau consists of only even harmonics instead of odd harmonics as shown in Fig.\ \ref{HHG_timedelay_phase} (b) since $\exp(i\Delta \phi_c)=-1$ in Eq.\ (\ref{XUV_symmetry}).
  Consequently, by changing the time delay of the XUV pulse, the hole transfer during the HHG process can be controlled. Moreover, the HHG process can probe this hole dynamics.
  The concept of coherent population transfer in a two-level system with a train of ultrashort laser pulses has previously been discussed for different systems \cite{Zhdanovich2008}. \par

\section{Conclusion} \label{Conclusion}
We studied HHG spectra produced by a driving IR field combined with XUV pulses. 
Our numerical method, based on a 1D-model atom using TDCIS calculation, shows good agreement with perturbative calculations based on SFA, with the production of a second plateau region that maps out the excursion times of electron trajectories driven by the intense laser field.   
This extended plateau originates from the recombination of a continuum electron with an inner shell hole that is generated by XUV-excitation after the tunneling process. The time delay between the XUV pulse and the IR field sows a control knob for the population transfer between the different ionic states. In the case of a single XUV attopulse and a few cycle IR pulse, analyzing the second plateau allows for extracting the temporal information of the HHG dynamics on a subcycle time-scale, such as ionization rate (intensity of the interference stripes), and ionization and recombination time of the classical trajectory (the start and end of the plateaus). In the region of overlapping attopulse and tunnel ionization time, the plateau as a function of time shows a positive slope that was attributed to modified effective tunneling time due to the XUV field. The tunneling time is a concept coming from the SFA and tunneling happens on the imaginary time axis. The perturbative XUV fields, when applied close to the tunnel ionization time, will change the imaginary tunneling time by introducing an additional delay dependent phase. This delay dependent phase is uniquely imprinted on the spectral slope of the second plateau and can be retrieved. 
Moreover, we showed that the second plateau can be increased by applying a combination of IR flat-top pulses and attosecond pulse trains. The increase of the plateau reveals out of a coherent accumulation of the inner-hole occupation in consequent half cycles. Our proposed XUV-assisted HHG spectroscopy would be realizable with existing table-top attosecond sources and will show more sight on attosecond electron dynamics in strong IR fields and hole dynamics in the residual ion.\par

\begin{acknowledgments}
J.M.D. acknowledges support from the Swedish Research Council, 
Grant No.  2013-344 and 2014-3724. 
\end{acknowledgments}

\appendix*
\section{THE FACTORIZATION OF THE TIME DEPENDENT DIPOLE IN SFA} 
If we apply the first excited state wave function of the 1D soft potential
\begin{equation}
V(x)=-\frac{Z_\text{eff}}{\sqrt{x^2+a^2}}
\end{equation}
to the state as the outer most electron wave function, this wave function has the asymptotic behavior
\begin{equation}
\lim_{|x| \rightarrow \infty} \langle x|2 \rangle	\approx xe^{-\sqrt{2 I_p}|x|}.
\end{equation}
We use the above approximation for the bound state wave function because the dominant pole in the bound-free dipole matrix element is determined by the asymptotic behavior \cite{Gribakin1997,Kuchiev1999}. This dipole matrix element can be approximated as
\begin{equation} \label{bc_dipole}
\text{d}_2(p+A(t_1))\approx\frac{1-8(p+A(t_1))^2}{((p+A(t_1))^2+2I_p)^2}
\end{equation}
When the canonical momentum is chosen as the stationary momentum $p_{(s)}(t,t_1)$ as shown in Eq.\ (\ref{momentum}), There is singularity in the denominator and it is exactly located in the saddle point $t_1=t_{1(s)}$ of $S_1$ (and $S_2$) even though $p=p_{(s)}(t,t_1)$ is chosen:
\begin{equation}
\frac{\partial S_1(t,t_1,p)}{\partial t_1}\Bigg|_{t_1=t_{1(s)}} =-\frac{[p+A(t_{1(s)})]^2}{2}-I_p=0.
\end{equation}
For a positive $I_p$, the solutions $t_1=t_{1(s)}$ of the above equation are moved to the complex plane from three step model ionization time $t_1=t_i$, which satisfies $p_{(s)}+A(t_{i})=0$.
Eq.\ (\ref{IRdipole}) can be approximated as
\begin{align}
& \int dt_1 \frac{1-8(p_{(s)}+A(t_1))^2}{[S_1'(t,t_1,p_{(s)})]^2}E_{\text{IR}}(t_1)a_\text{pr}(t,t_1)a_\text{rec}(t,t_1) \nonumber \\
\approx & \int dt_1\frac{[1-8(p_{(s)}+A(t_1))^2]E_{\text{IR}}(t_1)a_\text{pr}(t,t_1)a_\text{rec}(t,t_1)}{[S''(t,t_{1(s)},p_{(s)})]^2 (t_{1}-t_{1(s)})^2}
\end{align}
The stationary phase approximation with singularity at saddle point $t_1=t_{1(s)}$ should be modified \cite{Gribakin1997}:
\begin{equation}
\int_C dt_1 \frac{g(t_1)e^{-iS_1(t_{1})}}{(t_{1}-t_{1(s)})^2}=-\frac{\sqrt{2\pi}}{2}g(t_{1(s)})(S_1''(t_{1(s)}))^{\frac{1}{2}}e^{-iS_1(t_{1(s)})},
\end{equation}
where $g(t_1)$ is any slowly-varying function, and the variable $t$ and $p_{(s)}$ are neglected here. In the limit of strong field $U_p \gg I_p$, the complex time $t_{1(s)}$ is slightly shifted away from $t_i$. We can apply the Taylor expansion around $t_1=t_i$ to the $S_1(t_{1(s)})$ and $S_1''(t_{1(s)})$, the time dependent dipole moment is factorized as Eq.\ (\ref{IR_factorization}) and the detailed form of $a_{\text{ion}}$ is
    \begin{equation} \label{dipole_factorization}
    a_{\text{ion}}(t,t_{i})= \frac{C(I_p,p_{(s)}) E(t_{i})}{|E(t_{i})|^{\frac{3}{2}}}\exp \bigg[-\frac{(2I_p)^{3/2}}{3|E(t_{i})|}\bigg]
    \end{equation}
    where $C(I_p,p_{(s)})$ is a constant depending on $I_p$ and the canonical momentum $p_{(s)}$.
%

\end{document}